\documentclass[12pt,preprint]{aastex}
\usepackage{graphicx}
\usepackage{epsfig}

\shorttitle{{\it XMM} search for $X$-rays from MACHO-96-BLG-5}
\shortauthors{Nucita et al.}

\begin{document}


\title{{\it XMM} search for $X$-ray emission from the black hole candidate MACHO-96-BLG-5}


\author{A. A. Nucita\altaffilmark{1}, F. De Paolis\altaffilmark{1}, G. Ingrosso\altaffilmark{1} and D. Elia\altaffilmark{1}}
\affil{Dipartimento di Fisica, Universit\`a degli Studi di Lecce and
INFN, Sezione di Lecce, \\ CP 193, I-73100 Lecce, Italy}

\author{J. de Plaa \altaffilmark{2,3} and J. S. Kaastra\altaffilmark{2}}
\affil{SRON National Institute for Space Research, Sorbonnelaan 2,
3584 CA Utrecht, The Netherlands} \affil{Astronomical Institute,
Utrecht University, PO Box 80000, 3508 TA Utrecht, The Netherlands}


\begin{abstract}
MACHO-96-BLG-5 is a microlensing event observed towards the bulge of
the Galaxy with exceptionally long duration of $\sim 970$ days. The
microlensing parallax fit parameters were used to estimate the lens
mass $M=6^{+10}_{-3}$ M$_{\odot}$ and its distance $d$ which results
to be in the range $0.5$ kpc - $2$ kpc. The upper limit on the
absolute brightness for main-sequence stars of the same mass is less
than $1$ $L_{\odot}$ so that the lens is a good black hole
candidate. If it is so, the black hole would accrete by interstellar
medium thereby emitting in the $X$-ray band. Here, the analysis of
an {\it XMM} deep observation towards MACHO-96-BLG-5 lens position
is reported. Only an upper limit (at $99.8\%$ confidence level) to
the $X$-ray flux from the lens position of $9.10\times 10^{-15}$--
$1.45\times 10^{-14}$ erg cm$^{-2}$ s$^{-1}$ in the energy band
$0.2-10$ keV has been obtained from that deep observation allowing
to constrain the putative black hole accretion parameters.
\end{abstract}

\keywords{black hole
physics---stars:individual(MACHO-96-BLG-5)---X-ray:stars}

\section{Introduction}

Gravitational microlensing is nowadays a well established technique
to map both visible and dark matter throughout the Galaxy. The first
lines of sight to be explored have been those towards the Magellanic
Clouds and the bulge of the Galaxy, leading to the observation of
several hundreds of microlensing events (\citealt{alcock} and
\citealt{sumi}). The interpretation of the observations, although
debated and controversial, supports the existence in the galactic
halo of MACHOs (Massive Astrophysical Compact Halo Objects) with
mass of $\simeq 0.4~M_{\odot}$. Microlensing events towards the
bulge of the Galaxy are interpreted as due to lenses prevalently
belonging to the known stellar populations. However, several
long-duration events towards the bulge have significant
probabilities of being due to Black Hole (BH) lenses
(\citealt{bennett2002} and \citealt{mao2002}).

It is well known that the light curve of a typical microlensing
event depends on four parameters (e.g. the mass $M$ of the lens, the
observer-lens distance $d$, the observer-source distance $s$ and the
lens transverse velocity $v_{\bot}$), while only two quantities are
available from observations (the amplification at maximum and the
event duration). Therefore, it remains a degeneracy in microlensing
parameters, so that the lens mass can not be univocally determined,
but only a statistical information can be obtained from the analysis
of a large enough number of microlensing events.

The parameter degeneracy may be solved in a few parallax events,
which are events with duration long enough to make possible to
estimate $d$ and/or $v_{\bot}$. In this case, a fit procedure to the
event light curve allows to extract the value of the lens mass. This
is the case of at least 22 microlensing parallax events
\citep{poinsexter2005} observed by the MACHO collaboration towards
the bulge of the Galaxy. Among these events three are particularly
interesting: MACHO-96-BLG-5, MACHO-98-BLG-6 and MACHO-99-BLG-22. For
these three events, the lens mass has been estimated to be
$6_{-3}^{+10}$ M$_{\odot}$, $6_{-3}^{+7}$ M$_{\odot}$
\citep{bennett2002}, and  $11_{-6}^{+12}$ M$_{\odot}$
\citep{mao2002,agol2002}, respectively.

The observed upper limit on the absolute brightness of these lenses
is $<1$ $L_{\odot}$ so the lenses can not be stars but are most
likely BHs\footnote{ Poindexter et al. (\citeyear{poinsexter2005})
also concluded that MACHO-99-BLG-22 is a strong BH candidate
($78\%$), MACHO-96-BLG-5 is a marginal BH candidate ($37\%$) and
MACHO-98-BLG-6 is a weak BH candidate ($2.2\%$).}
\citep[see][]{bennett2002}. Stellar mass BHs there exist in the
Galaxy as a consequence of the evolution of massive stars, but all
the candidates known up to date are members of binary systems. On
the other hand, because of the shape of the microlensing light
curves, we expect that the lens is an isolated object, so that the
coordinates of these events give us the direction toward which
pointing an $X$-ray instrument and acquire information on the lens
nature.

Indeed, the putative BH lens may be accreting interstellar gas and
could be luminous in the X-ray band if it is within the thin gas
layer of the galactic disk \citep{mao2002}. A first estimate of the
X-ray flux in the 1-10 keV band leads to fluxes of $\simeq 10^{-15}$
erg cm$^{-2}$ s$^{-1}$ \citep{agolkam}.

In this paper we report on a $100$ ks {\it XMM} observation towards
the MACHO-96-BLG-5 position, in order to search for an $X$-ray
signature from the accretion process of the BH candidate.

The paper is structured as follows: In Sect. 2, we give a short
description of the gravitational lensing event MACHO-96-BLG-5 and in
Sect. 4 we discuss about the $X$-ray observations performed towards
the target by other $X$-ray telescopes. In Sect. 4 we report our
observational results. Finally, in Sect. 5 we address some
conclusion.

Before closing this section we would like to mention that detecting
isolated BHs is of great importance in astrophysics since this
should allow to validate or not the standard model for stellar
evolution and BH formation\footnote{Indeed, it could be that nature
has more difficulties in producing BHs than expected within the
standard stellar evolution model as recently emerged by CHANDRA
observation of the $X$-ray pulsar CXO J164710.2-455216, indicating
that pulsar progenitor had mass greater than about $40M_{\odot}$
\citep{muno}.}. Simple estimates indicate in fact that about $10^8$
BHs should be present throughout the Galaxy (se e.g. \citealt{st})
and this is in agrement with the chemical enrichment by supernovae
indicating that about $2\times 10^8$ BHs should have formed
\citep{samland}. The MACHO and OGLE groups claimed that three
isolated BHs have been detected by gravitational microlensing
indicating that at most $\simeq 5\times 10^8~(9M_{\odot}/M_{BH})$
BHs reside in the Milky Way disk (\citealt{agol2002},
\citealt{alcock2000}). However, the number of isolated BHs present
in the Galaxy should be confirmed by direct observations of their
$X$-ray or radio emission \citep{maccarone}.

\section{The gravitational lensing event MACHO-96-BLG-5}

Gravitational microlensing (i.e. gravitational deflection and
amplification) of electromagnetic waves is a well known phenomenon
predicted by the General Theory of Relativity.
%

In presence of a massive object (such as a MACHO, a star or a
stellar mass BH) close enough to the line of sight to a star
(source), an amplification of the received flux may be observed
depending on
%
the Einstein radius $R_E=\left[4GMd(s- d)/(c^2s)\right]^{1/2}$, $d$
and $s$ being the lens and source distances, respectively. Due to
the relative transverse velocity $v_{\bot}$ between the lens and the
line of sight to the source, the amplification is time-dependent and
the duration time scale of a microlensing event is given by $\Delta
T = 2R_E/v_{\bot}$.

For most observed microlensing events, the lens mass can only be
estimated very crudely based upon the observed time $\Delta T$. In
fact, since $s$ is usually known and only two observed quantities
(the maximum amplification and $\Delta T$) are available from
observations, it remains a degeneracy in microlensing parameters.

For long enough timescale events it is often possible to measure
parallax effects which make possible to estimate the lens distance
and/or its transverse velocity $v_{\bot}$. In this case, a fit to
the event light curve allows to extract the value of the lens mass.
For these events, the measurement of the projected speed of the lens
allows to relate the lens mass and the source distance by
\begin{equation}
M\simeq\frac{v_{\bot}^2\Delta T ^2
c^2}{16G}\frac{1-d/s}{d}~,\label{massa}
\end{equation}
from which it is possible to infer the lens mass once $d$ is known.

\begin{figure}
\epsscale{.80} \plotone{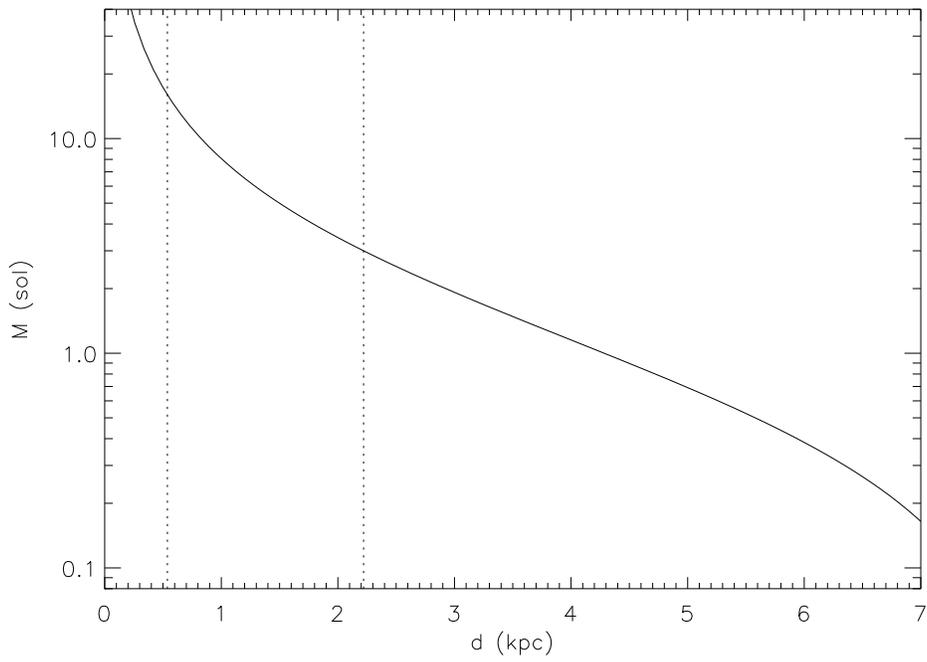} \caption{The parallax measurements
allow to correlate the lens mass $M$ to its distance $d$ from the
observer. According to the best fit parameters given in Table
\ref{table1}, the most plausible lens mass value is
$M=6^{+10}_{-3}$ M$_{\odot}$ corresponding to distances between
$\sim 0.5$ and $\sim 2$ kpc (dashed vertical
lines).}\label{figure1}
\end{figure}

Here, we focus on MACHO-96-BLG-5 event whose parameters are given in
Table \ref{table1} (\citealt{bennett2002}).
\begin{table}
\begin{center}
\caption{MACHO-96-BLG-5 parameters: event position in J2000
coordinates, duration, lens projected velocity and mass (Bennett et
al. 2002).}\label{table1}
\begin{tabular}{lllll}
\tableline\tableline RA(J2000)& DEC(J2000)& $\Delta T({\rm days})$ &$v_{\bot}({\rm km/s})$& $M(M_{\odot})$\\
$18:05:02.5$& $-27:42:17$&$970\pm 20$&$30.9\pm 1.3$&$6_{-3}^{+10}$\\
\tableline
\end{tabular}
\end{center}
\end{table}
MACHO-96-BLG-5 is a particularly long duration microlensing event so
that parallax measurement have been performed allowing to determine
the mass $M$ of the lens as a function of the distance $d$ (see
Figure \ref{figure1} where we plot the mass $M$ as given by eq.
(\ref{massa}) where we use the best fit parameters given in Table
\ref{table1}). The region between the two dashed vertical lines
corresponds to the lens mass best estimate $M=6^{+10}_{-3}$
M$_{\odot}$ corresponding to distances between $\sim 0.5$ kpc and
$\sim~2$~kpc.

\section{Previous $X$-ray observations}

An isolated BH may accrete the surrounding interstellar material as
a consequence of the deep gravitational potential. It is thus
expected that an $X$-ray emission, although weak (with respect to
neutron stars of the same mass), is present. This has motivated us
to search for an $X$-ray signature from the BH candidate in
MACHO-96-BLG-5 direction.

If the BH is moving with velocity $v$ with respect to the
interstellar medium it is expected to accrete at the Bondi-Hoyle
accretion rate given by (\citealt{st}, see also
\citealt{chisholsm2003} for a more recent discussion)
\begin{equation}
\dot{M}=4\pi \lambda
\frac{(GM)^2}{(a_{\infty}^2+v^2)^{3/2}}\rho_{\infty}
\end{equation}
where $\lambda$ is a constant of order unity, $a_{\infty}$ is the
sound velocity in the considered medium (which is in the range
$0.1$-$10$ km s$^{-1}$) and $\rho _{\infty}$ is the interstellar
medium density. Thus, the $X$-ray flux at Earth is
\begin{equation}
F_X=F_s \left(\frac{M}{M_{\odot}}\right)^2\left(\frac{1 {\rm
kpc}}{d}\right)^2~, \label{flusso1}
\end{equation}
where $F_s$ is a scale flux defined as
\begin{equation}
F_s =G^2c^2 \frac{\epsilon \rho_{\infty}}{(a_{\infty}^2+v^2)^{3/2}}
\left(\frac{1 M_{\odot}}{{\rm 1 kpc}}\right)^2~. \label{flusso2}
\end{equation}

Assuming $\rho_{\infty}\simeq 8.35\times10^{-25}$ g cm$^{-3}$
(corresponding to $n\simeq 0.5$ cm$^{-3}$ as expected for the
interstellar matter density) $\epsilon =0.1$ and $v\simeq 30$ km
s$^{-1}$ (as given by the microlensing event observation),
$F_s\simeq 4.45\times 10^{-15}$  erg cm$^{-2}$ s$^{-1}$.

Two observations have been carried out previously towards the
MACHO-96-BLG-5 coordinates by ROSAT and CHANDRA satellites.

\citet{maeda} have realized that the $9.7$ ks ROSAT observation in
the $0.1-2.4$ keV \citep{voges} made possible to detect $93$ and
$108$ photons from a circular source region with a radius of
$74\arcsec$ and an annular background region with inner and outer
radii of $74\arcsec$ and $120\arcsec$, respectively. Since the
expected background counts result to be $\simeq 0.00385$ counts
arcsec$^{-2}$, the number of background counts in the source circle
is $\simeq 66$. This corresponds to a source count number as high as
$\simeq 27$ corresponding to a signal-to-noise ratio of
$27/\sqrt{93}\simeq 2.8$. This signal-to-noise ratio is marginally
consistent with a source detection with a $2.8\sigma$ excess with a
flux of $\sim 4\times 10^{-14}$ erg cm$^{-2}$ s$^{-1}$ in the
$0.1-2.4$ keV band, or $\sim 1\times 10^{-13}$  erg cm$^{-2}$
s$^{-1}$ in the $0.3-8$ keV, where it has been assumed a photon
index of $2$ and an absorption column density of $3\times 10^{21}$
cm$^{-2}$ \citep{maeda}.

The $X$-ray signature from MACHO-96-BLG-5 has been also searched for
with a $10$ ks CHANDRA observation \citep{maeda}. In this case, the
CHANDRA ACIS-S image in the $0.3-8$ keV band has shown that not even
a single photon was detected within $1\arcsec$ of the target
position. In addition, $\simeq 21$ counts have been detected within
a circular region with $30\arcsec$ radius centered on the
MACHO-96-BLG-5 source. Hence, the background counts around the
target are $\simeq7\times10^{-3}$ counts arcsec$^{-2}$, which is
consistent with the nondetection of source photons within $1\arcsec$
from the MACHO-96-BLG-5 position. Using a simple Poissonian
distribution, Maeda et al. (\citeyear{maeda}) found an upper limit
of $4.6$ counts at $99\%$ confidence level and suggested that deeper
observations ($\gg 10$ ks) with CHANDRA or {\it XMM}-Newton may
detect, in principle, the weak $X$-ray signature from the lens of
MACHO-96-BLG-5 if it is really a BH. Assuming an interstellar column
density of $\simeq 3\times 10^{21}$ cm$^{-2}$, \citet{maeda} found
that the previous count estimate corresponds to fluxes of $5\times
10^{-15}$ erg cm$^{-2}$ s$^{-1}$ and $4\times 10^{-15}$ erg
cm$^{-2}$ s$^{-1}$, for photon index of $1.4$ and $2$, respectively.

\section{{\it XMM} Observation and Results}
A $100$ ks {\it XMM} observation towards the coordinates of
MACHO-96-BLG-5 has been made on October 2005 (Observation ID 30597)
with both MOS and PN cameras operating with thin filter mode.

\begin{figure}
\epsscale{.80} \plotone{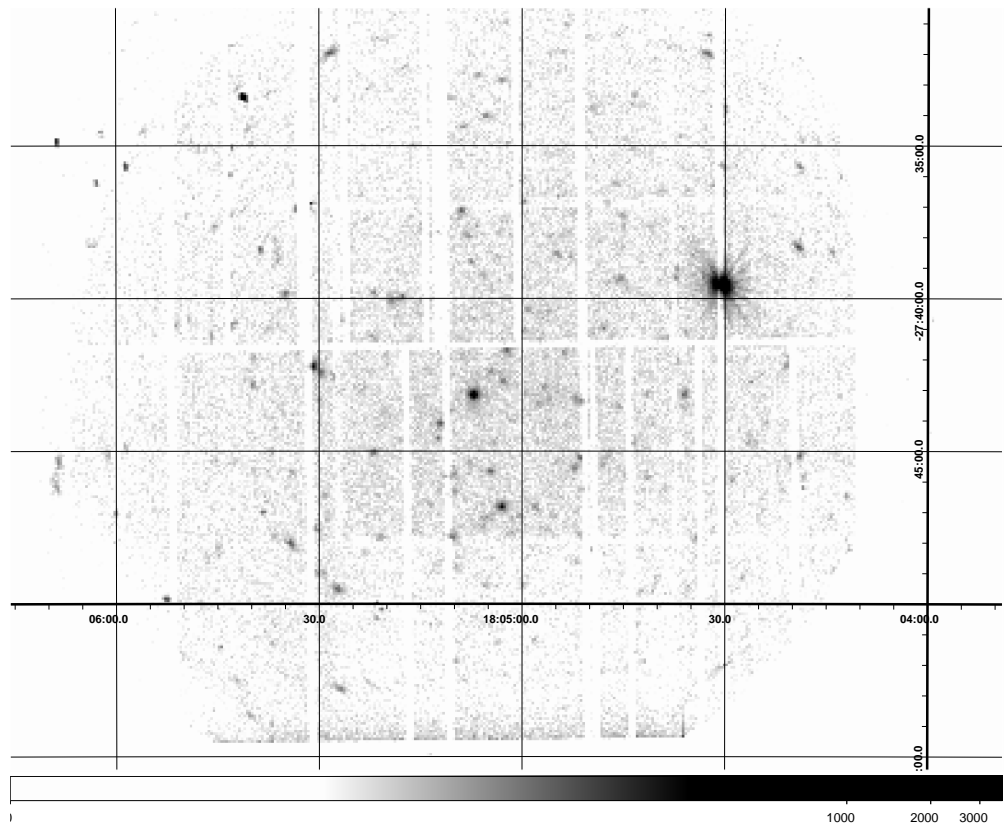} \caption{The deep {\it XMM} exposure
has shown the existence of a number of new $X$-ray sources which
will be investigated elsewhere.}\label{figure2}
\end{figure}

\begin{figure}
\epsscale{.80} \plotone{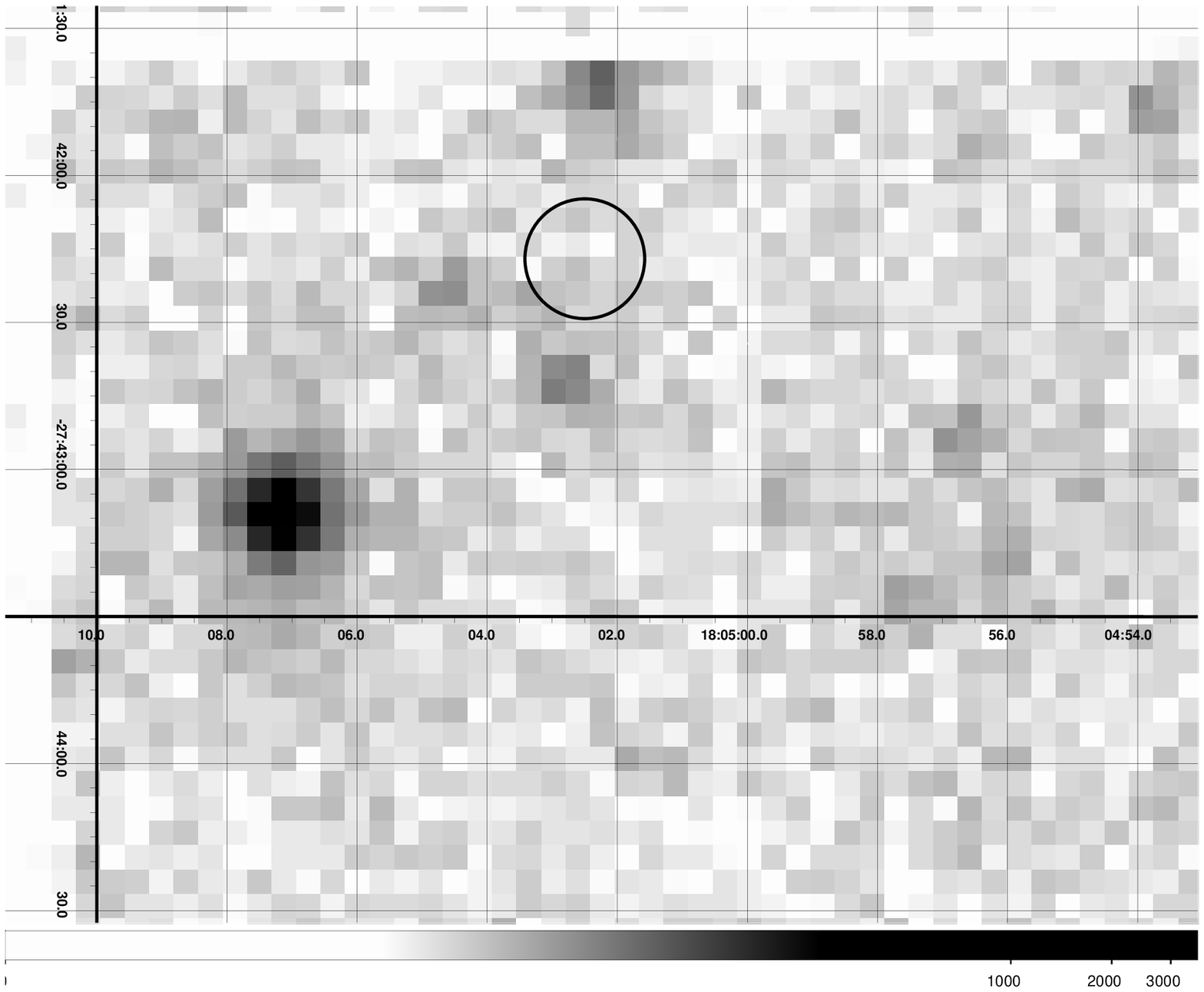} \caption{An enlargement of the {\it
XMM} exposure around the coordinates of the MACHO-96-BLG-5 target is
shown. In particular, the circle with radius $\simeq 15\arcsec$
(each pixel size is $\simeq 5\arcsec$) is centered on the target
coordinates given in Table \ref{table1}.}\label{figure3}
\end{figure}

The Epic Observation Data File (ODFs) where processed using the {\it
XMM}-Science Analysis System (SAS version $6.5.0$). With the latest
calibration constituent files available in 2006 May, we have
processed the raw data with the {\it emchain} and {\it epchain}
tools to generate proper event list files. We then only considered
events with patterns $0-12$ (resp. $0-4$) for the MOS (resp. PN)
instruments and we applied the filtering criterion XMMEA\_EM (resp.
FLAG$==0$) as recommended by the Science Operation Centre (SOC)
technical note XMM-PS-TN-43 v3.0. We have used the {\it evselect}
tool in order to extract light curves and images from the data. For
our pointing, we have rejected time periods affected by soft protons
flares which are evident, in the extracted light curves, as spikes.
For this purpose, we built light curves for the MOS and PN
instruments at energies
above $10$ keV (in particular in the energy band $10$ keV - $12$
keV) where the effect of soft protons flares is more evident. 
These data were recursively cleaned by removing all bins with counts
larger than $3\sigma$ from the mean. New mean and new deviation are
then found and the entire process has been repeated until a mean
count rate per bin is reached. This procedure, when applied to MOS
and PN data, allowed us to discard high background observing periods
on the basis of the derived thresholds of $22.5$ counts (or 0.225
counts sec$^{-1}$) for MOS 1, $26.0$ counts (or 0.26 counts
sec$^{-1}$) for MOS 2 and $49.04$ counts (or 0.490 counts
sec$^{-1}$) for PN, respectively. Hence, with the {\it tabgtigen}
tool, Good Time Intervals (GTIs) were obtained and used in order to
produce adequate $X$-ray event lists for each instruments.
The remaining good time intervals added up resulting in effective
exposures of $\simeq 95$ ks, $\simeq 98$ ks and $\simeq 67$ ks for
MOS 1, MOS 2 and PN, respectively.
In Figure \ref{figure2}, we show the full {\it XMM} field of view
(as a composite image of the MOS 1, MOS 2 and PN images in the
$0.2-10$ keV band) towards the MACHO-96-BLG-5 region. The deep
exposure has shown the existence of several new $X$-ray sources (not
seen in previous $X$-ray observations) which will be investigated
elsewhere. In Figure \ref{figure3}, we show an enlargement of the
previous figure around the coordinates of the MACHO-96-BLG-5 target.
In particular, the circle with radius $\simeq 15\arcsec$ (each pixel
size is $\simeq 5\arcsec$) is centered on the target coordinates
given in Table \ref{table1}. As it is evident from Figure
\ref{figure3}, if a $X$-ray source exists in the encircled region it
has to be extremely weak. In the following, we will quantify this
conclusion.

In order to confirm the absolute astrometry of our observations, we
searched for counterparts of the putative $X$-ray sources in the
Tycho-2 optical astrometric catalogue \citep{hog2000}, and we have
found a number of objects (whose coordinates are given in Table
\ref{tabella2}) corresponding to $X$-ray sources present in the {\it
XMM} field of view. 
\begin{deluxetable}{crrrrrrr}
\tabletypesize{\scriptsize} \tablecaption{A list of optical sources
(from Tycho-2 optical astrometric catalogue) whose coordinates
correspond to $X$-ray sources in our image is given.}
\tablewidth{400pt} \tablehead{\colhead{} &
\multicolumn{3}{c}{$\alpha$ (2000)} &
\multicolumn{3}{c}{$\delta$ (2000)} & V mag\\
\colhead{Source ID} &
\colhead{h}&\colhead{m}&\colhead{s}&\colhead{\degr}&\colhead{\arcmin}&\colhead{\arcsec}
\\} \startdata

       1 &       18 &        5 &       30.75 &      -27 &       42 &       12.790 & 11.460\\
       2 &       18 &        5 &       21.61 &      -27 &       29 &       40.50  & 11.163\\
       3 &       18 &        5 &       40.08 &      -27 &       53 &       38.20  & 10.160\\
       4 &       18 &        4 &       41.95&      -27 &       30 &       35.60  & 12.141\\
       5 &       18 &        5 &       3.55 &      -27 &       33 &       30.61 & 10.925\\
       6 &       18 &        5 &       31.94 &      -27 &       54 &       36.17 & 11.244\\

\enddata
\label{tabella2}
\end{deluxetable}

MOS 1, MOS 2 and PN images were produced in the $0.2-10$ keV band
with a binning that resulted in pixels of $5\arcsec$, close to the
full width at half maximum of the point spread function of the EPIC
instruments. None of the images showed an apparent excess in photons
at the location of the source. Images in narrower energy bands
across the $0.2-10$ keV range (i.e. 0.2-0.5 keV, 0.5-2 keV, 2-4.5
keV, 4.5-7 keV and 7-10 keV) were also produced but they did not
show an obvious source detection either.

We may also exclude the possibility that, as a consequence of the
proper motion, the BH candidate has moved away from its original
position. This possibility can be easily ruled out since, for a lens
distance in the range $0.5$ kpc--$2$ kpc (see Figure \ref{figure1}),
as deduced by the parallax measurement of the MACHO-96-BLG-5
microlensing event, and for a conservative relative velocity between
the lens and the observer of $\simeq 30$ km s$^{-1}$, the lens has
moved from its original position at most of $\simeq
0.015\arcsec-0.12\arcsec$ in a period of $10$ years. Hence, because
of the {\it XMM} pixel size quoted above, such a motion is
completely negligible in our case.

Next, we ran the {\it edetect\_chain} simultaneously on the three
images (MOS 1, MOS 2 and PN in the 0.2-10 keV band) produced. The
subtask {\it eboxdetect} was used with a detection threshold of {\it
likemin=8} to provide a complete source list as input for subsequent
tasks. No source was detected at the position of MACHO-96-BLG-5. As
suggested by the {\it User's Guide to the {\it XMM}-Newton Science
Analysis System} \citep{loiseau}, especially in cases where an
expected $X$-ray source is not detected by the usual source
detection tasks, we have used the {\it esenmap} command to generate
a sensitivity map roughly giving point source detection upper limit
(in units of counts $s^{-1}$). Performing this task on the MOS 1,
MOS 2 and PN images in the 0.2-10 keV band, we obtained the rates of
$3.1\times10^{-4}$ count $s^{-1}$, $3.0\times10^{-4}$ count $s^{-1}$
and $6.2\times10^{-4}$ count $s^{-1}$, respectively \footnote{As a
comparison, the weakest source detected with the automatic procedure
 has a cumulative count rate of $1.7\times10^{-3}$ count $s^{-1}$ ($2.2\times10^{-4}$ count $s^{-1}$ for MOS 1,
  $5.4\times10^{-4}$ count $s^{-1}$ for MOS 2 and $9.3\times10^{-4}$ count $s^{-1}$ for PN, respectively).}.


The $X$-ray rate from MACHO-96-BLG-5 in the $0.2-10$ keV band can be
determined in the following alternative way. We consider a source
extraction region around the target coordinates and several
background extraction regions sufficiently far from any bright point
source but on the same chip where our target is supposed to be. Let
us assume that the source and background extraction regions have
circular shape with the radii $R_s$ and $R_b$ and covering area
$A_s$ and $A_b$, respectively. For the source extraction region we
fixed $R_s=2.45$ pixels (or equivalently $12.5$ arcsec) so that it
is collected at least $65\%$ (resp. $60\%$) of the encircled energy
when the MOS (resp. PN) instrument is used. In the case of the
background extraction region a radius $R_b = 10$ pixels (resp. $R_b
= 6$ pixels) has been chosen for MOS 1 and MOS 2 (resp. PN).
Therefore, the covering areas are $A_s\simeq 475$ arcsec$^2$ and
$A_b\simeq 7775$ arcsec$^2$ for MOS 1 and MOS 2 and $A_b\simeq 2735$
arcsec$^2$ for PN, respectively. We also emphasize that the source
extraction region has been chosen to be smaller than the background
extraction region to avoid strong contamination by nearby sources.

Let $N_{s+b}$ and $N_b$ be the numbers of counts expected within the
source and background regions, respectively. We assume that within
the source circle, the observed count number depends on the
properties of both the source and background, while the counts
eventually observed within the background region are due to
background only. In this case the number of counts due to the source
is
\begin{equation}
N_s=N_{s+b}-\frac{A_s}{A_b}N_b~. \label{source}
\end{equation}
We tested different choices of background region positions, until we
were satisfied that the contaminations due to close sources were
properly subtracted. The number of counts detected within the circle
(with radius $R_s$) centered on the nominal source position were 60,
73 and 252 in the MOS 1, MOS 2 and PN cameras, respectively. Using
the tables in \citep{gelers}, this leads to single-sided $3\sigma$
(or $99.8\%$) upper limits of $87.1$, $102.5$ and $303.4$ on the
total counts. The expected background counts (after correcting for
exposure map differences and rescaling for the extraction area) are
56, 64 and 227 for MOS 1, MOS 2 and PN, respectively. Correcting for
the background, one has $31.1$, $38.5$ and $76.4$ counts
corresponding to the rates of $3.2\times10^{-4}$ count $s^{-1}$,
$3.9\times10^{-4}$ count $s^{-1}$ and $1.0\times10^{-3}$ count
$s^{-1}$, respectively.

We have also evaluated the expected 0.2-10 keV flux upper limit
using PIMMS. For this reason, we have assumed an interstellar
absorbtion column density of $3.81\times 10^{21}$ cm$^{-2}$ (as
estimated by the Nh Heasarc Tool) and a power law spectrum with
indices $\Gamma$ varying in the range 1.5 and 2.5. For the MOS 1
instrument the previously quoted count rates correspond to
unabsorbed fluxes of $9.10\times 10^{-15}$ erg cm$^{-2}$ s$^{-1}$
($\Gamma = 1.5$) and $1.20\times 10^{-14}$ erg cm$^{-2}$ s$^{-1}$
($\Gamma = 2.5$). For the MOS 2 instrument we get the unabsorbed
fluxes of $1.12\times 10^{-14}$ erg cm$^{-2}$ s$^{-1}$ ($\Gamma =
1.5$) and $1.45\times 10^{-14}$ erg cm$^{-2}$ s$^{-1}$ ($\Gamma =
2.5$). For EPIC-PN we find $1.03\times 10^{-14}$ erg cm$^{-2}$
s$^{-1}$ ($\Gamma = 1.5$) and $1.28\times 10^{-14}$ erg cm$^{-2}$
s$^{-1}$ ($\Gamma = 2.5$), respectively. Therefore, we can set an
upper limit for the flux from the putative source in the range
$9.10\times 10^{-15}$--$1.45\times 10^{-14}$ erg cm$^{-2}$ s$^{-1}$.

\section{Discussion and conclusions}

MACHO-96-BLG-5 is a microlensing observed event towards the bulge of
the Galaxy with the exceptionally long duration of $\sim 970$
days\footnote{We also mention that recently Poindexter et al.
(\citeyear{poinsexter2005}) presented the best fit solution for
geocentric timescale for the same microlensing event and found that
the event duration is $546 \pm 165$ days or $698 \pm 303$ days. This
analysis leads to the conclusion that the lens of MACHO-96-BLG-5 is
BH with a probability of about $37\%$.}. The measure of the parallax
angle \citep{bennett2002} made it possible to estimate both the lens
mass ($M=6^{+10}_{-3}$ M$_{\odot}$) and its distance $d$ from Earth
($0.5$--$2$ kpc).

Since the observed upper limits on the absolute brightness of
main-sequence stars for this lens is  $<1~L_{\odot}$, MACHO-96-BLG-5
is a BH candidate. If it is so, then the BH would accrete by
interstellar medium (via the Bondi-Hoyle mechanism) thereby emitting
$X$ rays. Here, we have reported about a deep {\it XMM} observation
in the $0.2-10$ keV band towards MACHO-96-BLG-5 lens position. The
analysis of the observations (see Section 4) did not show any
obvious source  and it is consistent with a non-detection of the
source. However, it is possible to estimate an upper limit on the
source luminosity, which turns out to be in the range $9.10\times
10^{-15}$-- $1.45\times 10^{-14}$ erg cm$^{-2}$ s$^{-1}$, depending
on the power law index $\Gamma$.

\begin{figure}
\epsscale{.80} \plotone{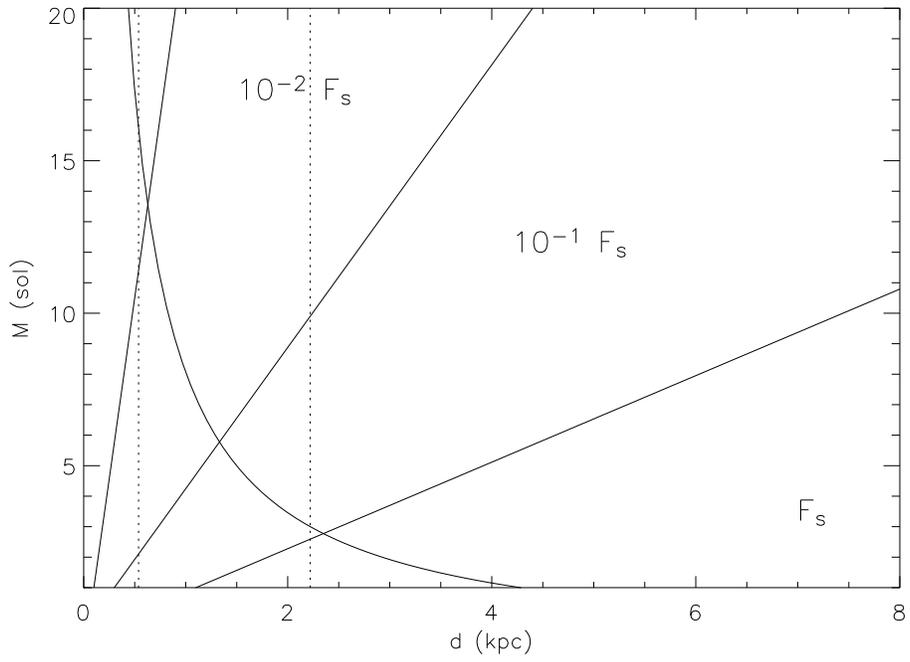} \caption{In the mass-distance plane
(see Figure \ref{figure1}), the three oblique lines define the
regions (below each line) where the $X$-ray flux due to the Bondi
accretion is below the upper limit flux of $9.10\times 10^{-15}$ erg
cm$^{-2}$ s$^{-1}$.}\label{par}
\end{figure}

This upper limit allows to constrain the accretion parameters of the
BH candidate from a parametric study involving eqs. (\ref{flusso1})
and (\ref{flusso2}). In the BH mass-distance plane given in Figure
\ref{par}, the curved line corresponds to the mass-distance relation
(also given in Figure \ref{figure1}). This means that our BH
candidate should lie on this curve (if its transverse velocity is
$\simeq 30$ km s$^{-1}$) and within the two dotted vertical lines.
The three oblique lines correspond to the $X$-ray upper limit of
$9.10\times 10^{-15}$ erg cm$^{-2}$ s$^{-1}$ for three different
values of the scale flux ($F_s$, $10^{-1}F_s$ and $10^{-2}F_s$).
Obviously, varying $F_s$, it means that the accretion parameters
($\epsilon$, $\rho_{\infty}$ and $v$) are changing. Thus, the region
below the line labeled $F_s$ corresponds to $F_X<9.10\times
10^{-15}$ erg cm$^{-2}$ s$^{-1}$. The same holds for the regions
below the other two lines. It is noticing that a BH accreting with
$\rho_{\infty}\simeq 8.35\times10^{-25}$ g cm$^{-3}$, $\epsilon
=0.1$ and $v\simeq 30$ km s$^{-1}$ (implying $F_s\simeq 4.45\times
10^{-15}$  erg cm$^{-2}$ s$^{-1}$), is inconsistent with the best
fit values (mass and distance) for the MACHO-96-BLG-5 lens
candidate. However, for a less efficient Bondi accretion ($\epsilon
\simeq 10^{-5}-10^{-3}$) the regions below the other two oblique
lines show that the agreement is recovered.

Finally, it could be interesting to estimate the BH radiative
efficiency $\eta=L_X/L_{Edd}$ with respect to the maximum allowed BH
accretion rate given by the Eddington luminosity $L_{Edd}\simeq
1.38\times10^{38}(M/M_{\odot})$ erg s$^{-1}$. By using the BH mass
and distance ranges, namely $3<M<16$ M$_{\odot}$ and $0.5<d<2$ kpc,
we can compare the Eddington luminosity with the $X$-ray luminosity
given by $L_X=F_X4\pi d^2$ expected for MACHO-96-BLG-5. This
procedure gives $\eta$ in the range $1.2\times 10^{-10}$ --
$2.7\times 10^{-9}$, in agreement with the efficiency estimates for
BH accretion in quiescent galaxies and ultra-low luminous AGNs for
which $\eta = 4\times 10^{-12}$-$~6\times 10^{-7}$ \citep{baganoff}.

%

\acknowledgments

This work has been partially supported by MIUR through PRIN 2004,
($2004020323\_ 004$). AAN is grateful to SRON for the research
facilities and the kind hospitality. We are also grateful to A. F.
Zakharov and B. M. T. Maiolo for interesting discussions. The
authors would like to thank the anonymous referee for suggestions
that improved the manuscript. {\it XMM-Newton} is an ESA science
mission with instruments and contributions funded by ESA member
states and NASA.

\end{document}